\documentclass[%
 reprint,
superscriptaddress,
 amsmath,amssymb,
 aps,
prc,
twocolumn,
]{revtex4}

\usepackage{graphicx}
\usepackage{dcolumn}
\usepackage{bm}

\begin{document}

\preprint{APS/123-QED}

\title{Charge-changing cross sections for $^{42\textrm{--}51}$Ca and effect of charged-particle evaporation induced by neutron removal reaction}

\author{M.~Tanaka}
\email{masaomi.tanaka@riken.jp}
\affiliation{RIKEN Nishina Center, Wako, Saitama 351-0198, Japan}
\affiliation{Department of Physics, Osaka University, Toyonaka, Osaka 560-0043, Japan}
\author{M.~Takechi}
\affiliation{Department of Physics, Niigata University, Ikarashi, Niigata 951-2181, Japan}
\author{A.~Homma}
\affiliation{Department of Physics, Niigata University, Ikarashi, Niigata 951-2181, Japan}
\author{A.~Prochazka}
\affiliation{GSI Helmholtzzentrum f\"ur Schwerionenforschung, 64291 Darmstadt, Germany}
\author{M.~Fukuda}
\affiliation{Department of Physics, Osaka University, Toyonaka, Osaka 560-0043, Japan}
\author{D.~Nishimura}
\affiliation{Department of Physics, Tokyo City University, Setagaya, Tokyo 158-8557, Japan}
\author{T.~Suzuki}
\affiliation{Department of Physics, Saitama University, Saitama 338-8570, Japan}
\author{T.~Moriguchi}
\affiliation{Institute of Physics, University of Tsukuba, Tsukuba, Ibaraki 305-8571, Japan}
\author{D.S.~Ahn}
\affiliation{RIKEN Nishina Center, Wako, Saitama 351-0198, Japan}
\author{A.~Aimaganbetov}
\affiliation{Institute of Nuclear Physics, 050032 Almaty, Kazakhstan}
\affiliation{L.N. Gumilyov Eurasian National University, 010008 Astana, Kazakhstan}
\author{M.~Amano}
\affiliation{Institute of Physics, University of Tsukuba, Tsukuba, Ibaraki 305-8571, Japan}
\author{H.~Arakawa}
\affiliation{Department of Physics, Saitama University, Saitama 338-8570, Japan}
\author{S.~Bagchi}
\altaffiliation[Present address: ]{Indian Institute of Technology (Indian School of Mines) Dhanbad, Jharkhand 826004, India.}
\affiliation{GSI Helmholtzzentrum f\"ur Schwerionenforschung, 64291 Darmstadt, Germany}
\affiliation{Astronomy and Physics Department, Saint Mary's University, Halifax, NS B3H 3C3, Canada}
\affiliation{Justus Liebig University, 35392 Giessen, Germany}
\author{K.-H.~Behr}
\affiliation{GSI Helmholtzzentrum f\"ur Schwerionenforschung, 64291 Darmstadt, Germany}
\author{N.~Burtebayev}
\affiliation{Institute of Nuclear Physics, 050032 Almaty, Kazakhstan}
\affiliation{Al-Farabi Kazakh National University, 050040 Almaty, Kazakhstan}
\author{K.~Chikaato}
\affiliation{Department of Physics, Niigata University, Ikarashi, Niigata 951-2181, Japan}
\author{H.~Du}
\affiliation{Department of Physics, Osaka University, Toyonaka, Osaka 560-0043, Japan}
\author{T.~Fujii}
\affiliation{Department of Physics, Saitama University, Saitama 338-8570, Japan}
\author{N.~Fukuda}
\affiliation{RIKEN Nishina Center, Wako, Saitama 351-0198, Japan}
\author{H.~Geissel}
\affiliation{GSI Helmholtzzentrum f\"ur Schwerionenforschung, 64291 Darmstadt, Germany}
\author{T.~Hori}
\affiliation{Department of Physics, Osaka University, Toyonaka, Osaka 560-0043, Japan}
\author{S.~Hoshino}
\affiliation{Department of Physics, Niigata University, Ikarashi, Niigata 951-2181, Japan}
\author{R.~Igosawa}
\affiliation{Department of Physics, Saitama University, Saitama 338-8570, Japan}
\author{A.~Ikeda}
\affiliation{Department of Physics, Niigata University, Ikarashi, Niigata 951-2181, Japan}
\author{N.~Inabe}
\affiliation{RIKEN Nishina Center, Wako, Saitama 351-0198, Japan}
\author{K.~Inomata}
\affiliation{Department of Physics, Saitama University, Saitama 338-8570, Japan}
\author{K.~Itahashi}
\affiliation{RIKEN Nishina Center, Wako, Saitama 351-0198, Japan}
\author{T.~Izumikawa}
\affiliation{Institute for Research Promotion, Niigata University, Niigata 950-8510, Japan}
\author{D.~Kamioka}
\affiliation{Institute of Physics, University of Tsukuba, Tsukuba, Ibaraki 305-8571, Japan}
\author{N.~Kanda}
\affiliation{Department of Physics, Niigata University, Ikarashi, Niigata 951-2181, Japan}
\author{I.~Kato}
\affiliation{Department of Physics, Saitama University, Saitama 338-8570, Japan}
\author{I.~Kenzhina}
\affiliation{Institute of Nuclear Physics, 050032 Almaty, Kazakhstan}
\affiliation{Al-Farabi Kazakh National University, 050040 Almaty, Kazakhstan}
\author{Z.~Korkulu}
\affiliation{RIKEN Nishina Center, Wako, Saitama 351-0198, Japan}
\author{Y.~Kuk}
\affiliation{Institute of Nuclear Physics, 050032 Almaty, Kazakhstan}
\affiliation{L.N. Gumilyov Eurasian National University, 010008 Astana, Kazakhstan}
\author{K.~Kusaka}
\affiliation{RIKEN Nishina Center, Wako, Saitama 351-0198, Japan}
\author{K.~Matsuta}
\affiliation{Department of Physics, Osaka University, Toyonaka, Osaka 560-0043, Japan}
\author{M.~Mihara}
\affiliation{Department of Physics, Osaka University, Toyonaka, Osaka 560-0043, Japan}
\author{E.~Miyata}
\affiliation{Department of Physics, Niigata University, Ikarashi, Niigata 951-2181, Japan}
\author{D.~Nagae}
\affiliation{RIKEN Nishina Center, Wako, Saitama 351-0198, Japan}
\affiliation{Research Center for Superheavy Elements, Kyushu University, Fukuoka 819-0395, Japan}
\author{S.~Nakamura}
\affiliation{Department of Physics, Osaka University, Toyonaka, Osaka 560-0043, Japan}
\author{M.~Nassurlla}
\affiliation{Institute of Nuclear Physics, 050032 Almaty, Kazakhstan}
\affiliation{Al-Farabi Kazakh National University, 050040 Almaty, Kazakhstan}
\author{K.~Nishimuro}
\affiliation{Department of Physics, Saitama University, Saitama 338-8570, Japan}
\author{K.~Nishizuka}
\affiliation{Department of Physics, Niigata University, Ikarashi, Niigata 951-2181, Japan}
\author{K.~Ohnishi}
\affiliation{Department of Physics, Osaka University, Toyonaka, Osaka 560-0043, Japan}
\author{M.~Ohtake}
\affiliation{RIKEN Nishina Center, Wako, Saitama 351-0198, Japan}
\author{T.~Ohtsubo}
\affiliation{Department of Physics, Niigata University, Ikarashi, Niigata 951-2181, Japan}
\author{S.~Omika}
\affiliation{Department of Physics, Saitama University, Saitama 338-8570, Japan}
\author{H.J.~Ong}
\affiliation{Research Center for Nuclear Physics, Osaka University, Ibaraki, Osaka 567-0047, Japan}
\author{A.~Ozawa}
\affiliation{Institute of Physics, University of Tsukuba, Tsukuba, Ibaraki 305-8571, Japan}
\author{H.~Sakurai}
\affiliation{RIKEN Nishina Center, Wako, Saitama 351-0198, Japan}
\affiliation{Department of Physics, University of Tokyo, Bunkyo-ku, Tokyo 113-0033, Japan}
\author{C.~Scheidenberger}
\affiliation{GSI Helmholtzzentrum f\"ur Schwerionenforschung, 64291 Darmstadt, Germany}
\author{Y.~Shimizu}
\affiliation{RIKEN Nishina Center, Wako, Saitama 351-0198, Japan}
\author{T.~Sugihara}
\affiliation{Department of Physics, Osaka University, Toyonaka, Osaka 560-0043, Japan}
\author{T.~Sumikama}
\affiliation{RIKEN Nishina Center, Wako, Saitama 351-0198, Japan}
\author{H.~Suzuki}
\affiliation{RIKEN Nishina Center, Wako, Saitama 351-0198, Japan}
\author{S.~Suzuki}
\affiliation{Institute of Physics, University of Tsukuba, Tsukuba, Ibaraki 305-8571, Japan}
\author{H.~Takeda}
\affiliation{RIKEN Nishina Center, Wako, Saitama 351-0198, Japan}
\author{Y.~Tanaka}
\affiliation{Department of Physics, Osaka University, Toyonaka, Osaka 560-0043, Japan}
\author{Y.K.~Tanaka}
\affiliation{GSI Helmholtzzentrum f\"ur Schwerionenforschung, 64291 Darmstadt, Germany}
\author{I.~Tanihata}
\affiliation{Research Center for Nuclear Physics, Osaka University, Ibaraki, Osaka 567-0047, Japan}
\affiliation{School of Physics and Nuclear Energy Engineering, Beihang University, 100191 Beijing, China}
\author{T.~Wada}
\affiliation{Department of Physics, Niigata University, Ikarashi, Niigata 951-2181, Japan}
\author{K.~Wakayama}
\affiliation{Department of Physics, Saitama University, Saitama 338-8570, Japan}
\author{S.~Yagi}
\affiliation{Department of Physics, Osaka University, Toyonaka, Osaka 560-0043, Japan}
\author{T.~Yamaguchi}
\affiliation{Department of Physics, Saitama University, Saitama 338-8570, Japan}
\author{R.~Yanagihara}
\affiliation{Department of Physics, Osaka University, Toyonaka, Osaka 560-0043, Japan}
\author{Y.~Yanagisawa}
\affiliation{RIKEN Nishina Center, Wako, Saitama 351-0198, Japan}
\author{K.~Yoshida}
\affiliation{RIKEN Nishina Center, Wako, Saitama 351-0198, Japan}
\author{T.K.~Zholdybayev}
\affiliation{Institute of Nuclear Physics, 050032 Almaty, Kazakhstan}
\affiliation{Al-Farabi Kazakh National University, 050040 Almaty, Kazakhstan}

\date{\today}

\begin{abstract}
Charge-changing cross sections $\sigma_\mathrm{CC}$ for $^{42\textrm{--}51}$Ca on a carbon target at around 280~MeV/nucleon have been measured. The measured $\sigma_\mathrm{CC}$ values differ significantly from the previously developed calculations based on the Glauber model. However, through introduction of the charged-particle evaporation effect induced by the neutron-removal reaction in addition to the Glauber-model calculation, experimental $\sigma_\mathrm{CC}$ values on $^{12}$C at around 300~MeV/nucleon for nuclides from C to Fe isotopes are all reproduced with approximately 1\% accuracy. This proposed model systematically reproduces $\sigma_\mathrm{CC}$ data without phenomenological corrections, and can also explain experimental $\sigma_\mathrm{CC}$ values obtained in other energy regions.
\end{abstract}


\maketitle

\section{INTRODUCTION}
The point-proton radius $r_\mathrm{p}$ of the atomic nucleus, usually defined as the root mean square~(RMS) radius of the point-proton density distribution, is one of the key quantities to study nuclear structures.
Point-proton or charge radii have been measured using the electron elastic scattering, muonic X-ray, and optical isotope shift~(IS) methods~\cite{AN13}. Among these, the IS measurement is generally regarded as the only way to extract the $r_\mathrm{p}$ of unstable nuclei. Systematic $r_\mathrm{p}$ investigations have helped clarify exotic phenomena, such as the neutron-halo structure~\cite{WA04,SA06} and the dramatic enhancement of nuclear radii beyond the magic numbers~\cite{GA16,GO19}. The neutron-skin thickness can be also extracted from $r_\mathrm{p}$ combined with the matter radius~$r_\mathrm{m}$~\cite{SU95,TA20}. The pursuit of neutron-skin thickness has attracted significant attention particularly for very neutron-rich nuclei, to elucidate the density-dependent parameter of the symmetry energy term in the nuclear-matter equation of state~\cite{BR00}. However, the IS method is inapplicable to unstable nuclei far from the stability line, or to certain elements, because of beam production difficulties.

Alternative $r_\mathrm{p}$ determination methods for unstable nuclei have been proposed, based on reaction cross sections on proton and carbon targets~\cite{NI10,MO13,HO14}, proton elastic scattering at double energies~\cite{SAK17}, and electron elastic scattering under trapping in a storage ring~\cite{SU09,WA13,TS17}. The charge-changing cross section~$\sigma_\mathrm{CC}$, defined as the atomic-number-changing total cross sections, can potentially also be used to derive $r_\mathrm{p}$. Similar to the interaction cross section~$\sigma_\mathrm{I}$ or reaction cross section~$\sigma_\mathrm{R}$, which are sensitive to $r_\mathrm{m}$, $\sigma_\mathrm{CC}$ is used to probe $r_\mathrm{p}$~\cite{CH00,OZ01}. Furthermore, $\sigma_\mathrm{CC}$ can be measured even with low-intensity heavy-ion beams (e.g., a few particles per second~(pps)). Therefore, this method is a potential tool to study the $r_\mathrm{p}$ of a very neutron-rich nucleus.

To date, $\sigma_\mathrm{CC}$ measurement has been utilized to derive the $r_\mathrm{p}$ of light-mass nuclei~\cite{YA10,YA11,YA13,YA14,OZ14,SA17,TE14,ES14,KA16,BA19,TR16,TR18,ZH20}. Several methods based on the Glauber model have been proposed to describe the relationship between $\sigma_\mathrm{CC}$ and $r_\mathrm{p}$. For example, Yamaguchi~\emph{et al.} previously introduced an empirical scaling factor for the Glauber-model calculation to explain experimental $\sigma_\mathrm{CC}$ data for $^{28}$Si on a carbon target at intermediate energies of 100--600~MeV/nucleon~\cite{YA10}. This phenomenological model universally explained $\sigma_\mathrm{CC}$ data at 300~MeV/nucleon for light-mass nuclei over a wide range of mass-to-atomic-number ratios~$A/Z$~\cite{YA11}. However, some $\sigma_\mathrm{CC}$ data for medium-mass nuclides around calcium deviate from the above phenomenological-model calculation~\cite{YA13,YA14}. In contrast, $\sigma_\mathrm{CC}$ data for nuclei up to nitrogen at $\sim$900~MeV/nucleon have been explained by a Glauber-model calculation without the above scaling factor~\cite{TE14,ES14,KA16,BA19}. To explain $\sigma_\mathrm{CC}$ for $^{12}$C on $^{12}$C at 10--2100~MeV/nucleon, Tran~\emph{et al.} tuned the slope parameter of the proton--neutron elastic differential cross section, $\beta_{ij}$, which is one of the parameters in the Glauber-model calculation~\cite{TR16}. However, their model underestimated the $\sigma_\mathrm{CC}$ data for $^{12}$C on $^{12}$C at 200--400~MeV/nucleon. Although several theoretical studies have investigated this problem~\cite{SU16,FA19,AB20}, a consistent universal model for $r_\mathrm{p}$ derivation from $\sigma_\mathrm{CC}$ has not been established. Crucially, the mechanism underlying the discrepancy between the experimental data and the Glauber-model calculation remains unknown.


In this study, we report measurement of $\sigma_\mathrm{CC}$ on a carbon target at around 280~MeV/nucleon for $^{42\textrm{--}51}$Ca, for which $r_\mathrm{p}$ was previously measured via the IS method~\cite{GA16}. The obtained $\sigma_\mathrm{CC}$ results show a significant decrease with increases in the neutron number, a trend that differs from those in light-mass isotopic chains. Based on a comparison between the experimental results and Glauber-model calculations (taking the known $r_\mathrm{p}$ as input values), the reaction dynamics in the charge-changing process were investigated. The previously developed calculations could not explain the trends of experimental $\sigma_\mathrm{CC}$ values, especially in the medium-mass region. Finally, we propose a model that systematically reproduces $\sigma_\mathrm{CC}$ data in wide mass and energy regions without any empirical corrections by considering the charged-particle evaporation effect induced by the neutron-removal reactions.

\section{EXPERIMENT AND ANALYSIS}
\subsection{Experiment}
The experiment was conducted at the RI Beam Factory~(RIBF), operated by the RIKEN Nishina Center, and the Center for Nuclear Study, University of Tokyo. A 345-MeV/nucleon $^{238}$U primary beam and a rotating beryllium production target were used to produce $^{42\textrm{--}51}$Ca secondary beams. The secondary beams produced at the F0 focal plane were roughly purified in the first stage of the BigRIPS fragment separator~\cite{KU12} between the F0 and F3 focal planes. Then, $\sigma_\textrm{CC}$ was measured between the F3 and F7 focal planes. Owing to the large acceptance of BigRIPS, experimental data were acquired for three or four Ca isotopes simultaneously in a single BigRIPS setting.

The transmission method was applied to measure $\sigma_\mathrm{CC}$~\cite{TA13}, where
\begin{equation}
 \sigma_\mathrm{CC}=-\cfrac{1}{N_\mathrm{t}}\ln\left(\cfrac{\gamma}{\gamma_\mathrm{0}}\right),
\end{equation}
with $N_\mathrm{t}$ being the number of target nuclei per unit area and $\gamma$ and $\gamma_0$ the non-reaction rates with and without the reaction target, respectively. Note that, for $\sigma_\mathrm{CC}$ measurement, outgoing particles with the same~$Z$ as the incoming ones correspond to the non-reaction events. A wedge-shaped carbon target with an angle of 9.61~mrad was placed at the F5 momentum-dispersive focal plane to maintain the achromatic property of the F7 focal plane. The central-point thickness of the target was 1.803(3)~g/cm$^2$. The $\sigma_\mathrm{CC}$ value with the wedge-shaped target was obtained from the values at each position, i.e., $\sigma_\mathrm{CC}(X)$, weighted with the incident-particle distribution on the target $N_\mathrm{in}(X)$, where $X$ is the position along the momentum-dispersive (horizontal) direction perpendicular to the beam axis~$z$. The target-thickness profile $d(X)$ was measured with 0.15\% accuracy or higher. The mean energies in the reaction target, $E_\mathrm{mean}$, at the weighted mean position of $N_\mathrm{in}(X)$ are listed in Table~\ref{tab1}. The average $E_\mathrm{mean}$ of the $^{42\textrm{--}51}$Ca data was approximately 280~MeV/nucleon. Note that the experimental $\sigma_\mathrm{CC}$ data reported herein were obtained simultaneously with $\sigma_\mathrm{I}$ data for $^{42\textrm{--}51}$Ca~\cite{TA20}.

\begin{figure}[b]
\resizebox{0.5\textwidth}{!}{\includegraphics{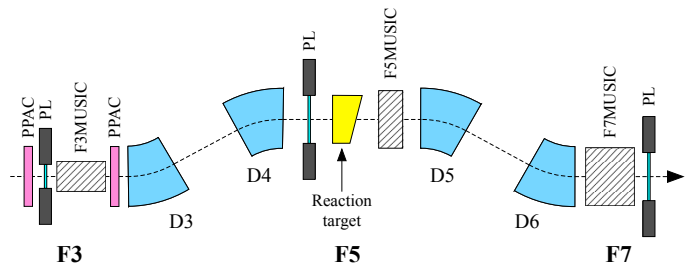}}
\caption{Schematic view of experimental setup.}
\label{setup}
\end{figure}
Figure~\ref{setup} shows the experimental setup between the F3 and F7 focal planes of BigRIPS. To derive the non-reaction rate, the incoming and non-reacting outgoing particles were counted before and after the reaction target, respectively. For particle identification~(PID) before the reaction target, the mass-to-charge ratio~$A/Q$ and $Z$ of the incoming particle were identified in event-by-event mode via the $B\rho$--TOF--${\Delta}E$ method between the F3 and F5 focal planes, where $B\rho$, TOF, and $\Delta{E}$ represent the magnetic rigidity, time of flight, and energy loss, respectively. Here, $B\rho$ was determined from the dipole-magnet magnetic field data together with the beam-ray tracking using parallel plate avalanche counters (PPACs) at F3 and a plastic scintillation counter~(PL) at F5, which was sensitive to~$X$. The TOF was measured by PLs installed at F3 and F5. A multi-sampling ionization chamber~(MUSIC) at F3~(F3MUSIC) was used to measure $\Delta{E}$.

In the downstream side of the reaction target, i.e., between F5 and F7, $Z$ was identified from the $\Delta{E}$ measured by two MUSICs installed at F5 and F7~(F5MUSIC and F7MUSIC, respectively). F5MUSIC is a large acceptance specification~(240~mm $\times$ 150~mm area and 200~mm length), whereas F7MUSIC is a high resolution specification~(240~mm$\phi$ area and 480~mm length). Between the F5 and F7 focal planes, BigRIPS was tuned to transport particles that did not change both $A$ and $Z$ at the reaction target. Therefore, only non-nuclide-changing and one-neutron-removal events of Ca isotopes ($^{42}$Ca and $^{41}$Ca in Fig.~\ref{pid}(b)) were transported to the F7 focal plane, owing to the $\pm3\%$ momentum acceptance of BigRIPS. The $Z$ resolution of F7MUSIC was much higher than that of F5MUSIC. For clear identification and reliable counting of the non-reacting particles, the non-nuclide-changing and one-neutron-removal particles of Ca isotopes were identified by using F7MUSIC. The other Ca-isotope particles were identified from the $\Delta{E}$ data of F5MUSIC only. To ensure the full acceptance of non-reacting particles after the reaction target, the position, angle, and momentum information from upstream detectors was constrained. This constraint was optimized for nuclides of interest.

\subsection{Data analysis and results}
\begin{figure}[t]
\resizebox{0.5\textwidth}{!}{\includegraphics{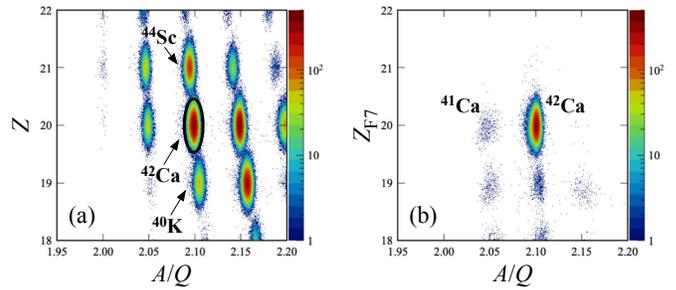}}
\caption{Particle-identification plots of (a) cocktail beam for the $^{42}$Ca setting before the reaction target and (b) particles transported to the F7 focal plane, with incoming $^{42}$Ca selection indicated by the ellipse in~(a). In~(b), $Z$ was determined from F7MUSIC $\Delta{E}$. The appropriate beam-emittance constraint was adopted for both plots. The $^{48}$Ca plots are presented in Ref.~\cite{TA20}.}
\label{pid}
\end{figure}
\begin{figure}[t]
\resizebox{0.35\textwidth}{!}{\includegraphics{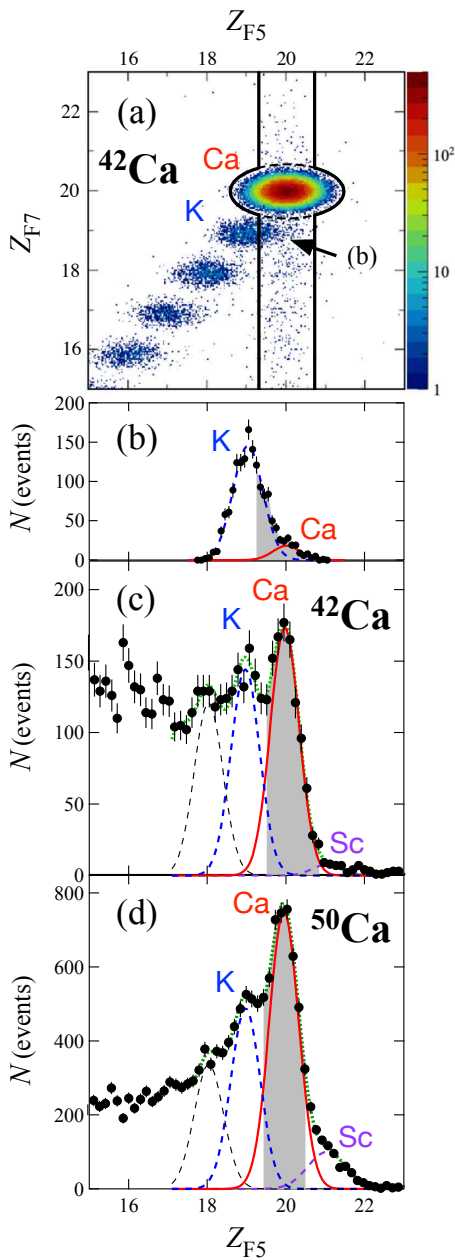}}
\caption{(a) Identification of $Z$ after the reaction target with selection of incoming $^{42}$Ca in correlation between $Z_\mathrm{F5}$ and $Z_\mathrm{F7}$. The events inside the black lines were counted. (b) $Z_\mathrm{F5}$ histogram for $Z_\mathrm{F7}=19$ events indicated by arrow in plot~(a). The shaded region corresponds to the area enclosed by the black line in plot (a). (c, d) $Z_\mathrm{F5}$ histograms for events that were not transported to F7 in cases of $^{42}$Ca and $^{50}$Ca, respectively. The dotted green line represents the fitting result. The red solid and dashed lines indicate the compositions.}
\label{zzpid}
\end{figure}
Figure~\ref{pid}(a) shows the typical PID plot for the beam before the reaction target for $^{42}$Ca, as an example. The nuclides were separated with 19.2$\sigma$ and 6.5$\sigma$ resolutions on $A/Q$ and $Z$, respectively. The $^{42}$Ca incident particles were selected by an elliptical gate with a width of $3.5\sigma$ of each axis. Contamination from neighboring nuclides, i.e., $^{44}$Sc and $^{40}$K, was excluded through additional selection in the correlation between the $\Delta{E}$ in F3MUSIC and those in the F3 and F5 PLs. Finally, these contaminants had effects far lower than 0.1\% on $\sigma_\mathrm{CC}$.

Figure~\ref{zzpid} shows the $Z$ identification plot for the incoming Ca isotopes after the reaction target. The events transported to the F7 focal plane were identified from the correlation plot between the atomic numbers determined by F5MUSIC and F7MUSIC, i.e., $Z_\mathrm{F5}$ and $Z_\mathrm{F7}$, respectively~(Fig.~\ref{zzpid}(a)). The peak separations of $Z_\mathrm{F5}$ and $Z_\mathrm{F7}$ were 3.0$\sigma$ and 6.3$\sigma$, respectively. The events within the black lines in Fig.~\ref{zzpid}(a) were counted as non-reacting particles. Here, the widths of elliptical and vertical-lines regions are $4\sigma$ and $2\sigma$, respectively. The K-isotope contaminants indicated by the arrow were subtracted by fitting the $Z_\mathrm{F5}$ histogram shown in Fig.~\ref{zzpid}(b).
The particles that were not transported to F7 were identified in the $Z_\mathrm{F5}$ histograms shown in Figs.~\ref{zzpid}(c, d). Although the $Z_\mathrm{F5}$ resolution was insufficient for complete peak separation, fitting was achieved with the help of $Z_\mathrm{F7}$. In this fit, the position and width of each element in $Z_\mathrm{F5}$ were constrained using data on the peaks tagged by $Z_\mathrm{F7}$ in Fig.~\ref{zzpid}(a). The events in the shaded areas, where the Ca events are dominant, were regarded as non-reacting particles. The Ca events outside this region and neighboring-element contaminants were corrected based on the fitted distributions (the red solid and dashed lines in Figs.~\ref{zzpid}(c, d)). The event-counting uncertainty in the identification only on $Z_\mathrm{F5}$ was typically 7\%. The ratio of the number of non-reaction events identified in Fig.~\ref{zzpid}(c, d), $N_\mathrm{F5}$, to that identified in Fig.~\ref{zzpid}(a), $N_\mathrm{F7}$, was larger for a neutron-rich nucleus. For example, the $N_\mathrm{F5}/N_\mathrm{F7}$ values for $^{42}$Ca and $^{50}$Ca were 0.0079(4) and 0.0285(20), respectively.
\begin{table}
\caption{Measured charge-changing cross sections $\sigma_\textrm{CC}$ for $^{42\textrm{--}51}$Ca on $^{12}$C target. The mean energies in the reaction target are listed in the second column. The first and second parentheses in the third column contain the statistical and systematic uncertainties, respectively.}
\begin{ruledtabular}
\begin{tabular}{ccc}
 & $E_\mathrm{mean}$ & $\sigma_\mathrm{CC}$\\
Nuclide & (MeV/nucleon) & (mb)\\ \hline
$^{42}$Ca & 297 & \multicolumn{1}{l}{1378(11)(6)}   \\
$^{43}$Ca & 284 & \multicolumn{1}{l}{1352(9)(7)}    \\
$^{44}$Ca & 270 & \multicolumn{1}{l}{1351(10)(10)}  \\
$^{45}$Ca & 302 & \multicolumn{1}{l}{1291(6)(10)}   \\
$^{46}$Ca & 290 & \multicolumn{1}{l}{1300(8)(15)}   \\
$^{47}$Ca & 277 & \multicolumn{1}{l}{1283(14)(14)} \\
$^{48}$Ca & 300 & \multicolumn{1}{l}{1259(14)(16)} \\
$^{49}$Ca & 291 & \multicolumn{1}{l}{1280(8)(18)} \\
$^{50}$Ca & 283 & \multicolumn{1}{l}{1297(11)(23)} \\
$^{51}$Ca & 271 & \multicolumn{1}{l}{1319(33)(28)} \\
\end{tabular}
\end{ruledtabular}
\label{tab1}
\end{table}

The experimental data without the reaction target for the respective isotopes were similarly analyzed. For example, $\gamma$ and $\gamma_0$ of $^{42}$Ca were 0.8744(8) and 0.9914(4), respectively. Table~\ref{tab1} summarizes the obtained $\sigma_\textrm{CC}$ for $^{42\textrm{--}51}$Ca on $^{12}$C. The statistical uncertainties (first parentheses) were typically less than 1.0\%. The main source of systematic uncertainty~(second parentheses) was the accuracy of $N_\mathrm{F5}$. As mentioned above, $N_\mathrm{F5}/N_\mathrm{F7}$ increases with increases in the neutron number of the nuclide of interest. Therefore, the total uncertainty is governed by a systematic uncertainty, especially in neutron-rich isotopes. Below, we treat the square root of the sum of these two uncertainties as the total uncertainty.

\begin{figure}[t]
\resizebox{0.5\textwidth}{!}{\includegraphics{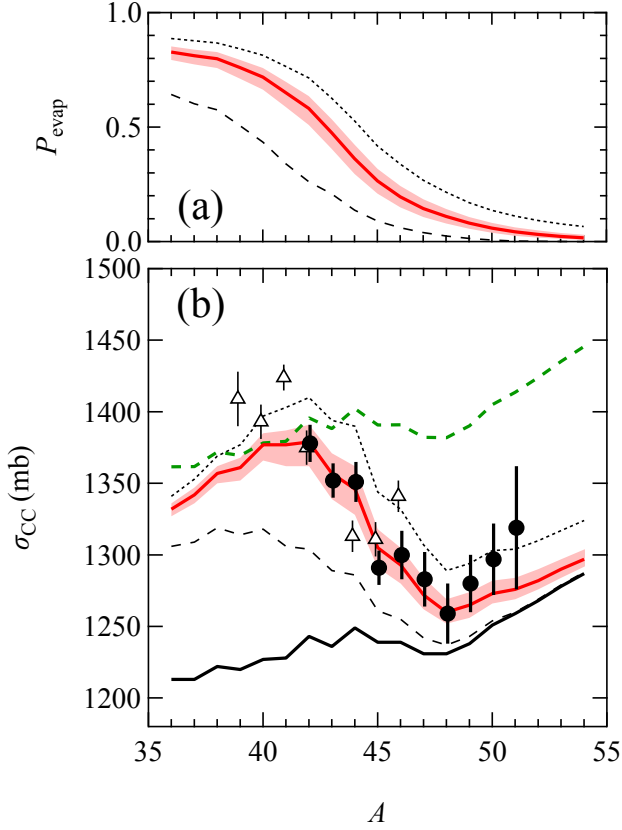}}
\caption{(a) $A$ dependence of $P_\mathrm{evap}$ values (Eq.~(\ref{math_pevap0})) with $E_\mathrm{max}=20$~MeV~(thin black dashed line), 45(8)~MeV~(red solid line), and 70~MeV~(thin black dotted line). (b) $A$ dependence of $\sigma_\mathrm{CC}$ for Ca isotopes on $^{12}$C. The present $^{42\textrm{--}51}$Ca results and existing data~\cite{YA13,YA14} are indicated by closed circles and open triangles, respectively. The black solid line and green dashed line represent the $\sigma_\mathrm{CC}$ values from the Glauber(ZROLA) calculation~(Eq.~(\ref{cccs_bare})) and the Glauber-model(ZROLA) calculation with the correction factor~\cite{YA10} (Eq.~(\ref{cccs_corr})), respectively. The thin black dashed line, red solid line, and thin black dotted line show the $\sigma_\mathrm{CC}$ calculations from the Glauber(CE) model~(Eq.~(\ref{cccs_ce})) with $E_\mathrm{max}=20$~MeV, 45(8)~MeV, and 70~MeV, respectively.}
\label{cccs_exp}
\end{figure}

\section{DISCUSSION}
Figure~\ref{cccs_exp}(b) shows the present $\sigma_\mathrm{CC}$ results for Ca isotopes as a function of mass number~$A$. The existing data are also shown for comparison~\cite{YA13,YA14}. The minimum energy of the present study and the energy of existing data were $E=271$ and 300~MeV/nucleon, respectively. A negligibly small difference of 0.2\% in $\sigma_\mathrm{CC}$ due to its energy dependence was estimated between these energies. Therefore, this slight difference was ignored and the calculation was performed for 280~MeV/nucleon. The present and existing data are consistent in $^{42,44\textrm{--}46}$Ca. Upon extension of the experimental data to the neutron-rich region, $\sigma_\mathrm{CC}$ decreased with increasing mass or neutron number. This $A$ dependence significantly differs from the experimental $\sigma_\mathrm{CC}$ trend in the light-mass region~\cite{CH00,TE14,ES14,KA16,BA19,TR16}, which shows rather flat dependence on $A$. To understand the trend of the experimental $\sigma_\mathrm{CC}$ of Ca isotopes, we first performed the Glauber-model calculation.

\subsection{Glauber-model calculation}
\begin{figure}[b]
\resizebox{0.5\textwidth}{!}{\includegraphics{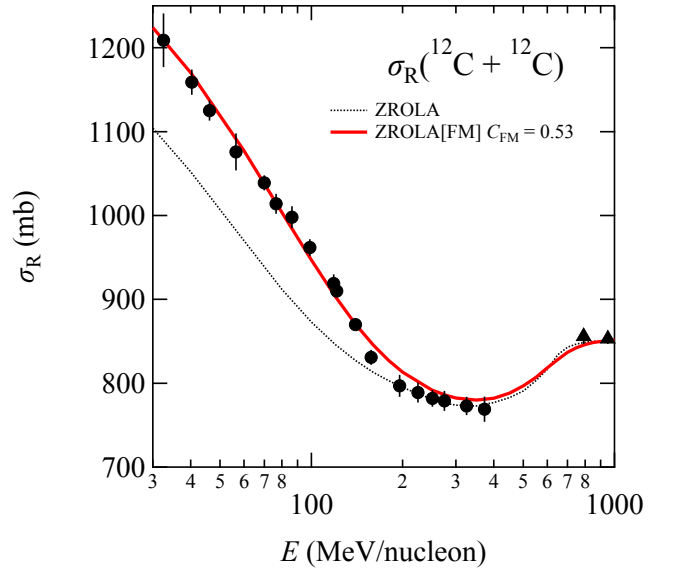}}
\caption{Energy dependence of $\sigma_\mathrm{R}$ for $^{12}$C on $^{12}$C~\cite{OZ01,OZ01b,TA09}. The red solid and black dotted lines indicate the present ZROLA calculation with the best-fit parameter for the Fermi motion, $C_\mathrm{FM}=0.53$, and the conventional ZROLA calculation without the Fermi motion effect on $\sigma_{ij}$, respectively.}
\label{RCS12C12C}
\end{figure}

There are several types of Glauber model calculations for $\sigma_\mathrm{R}$ and $\sigma_\mathrm{CC}$ that incorporate various effects such as multiple scattering~\cite{AB00a}, energy-dependent range parameters~\cite{HO07,TR16}, and Fermi motion~\cite{TA09}. Here, to describe both the $\sigma_\mathrm{R}$ and $\sigma_\mathrm{CC}$ in the same framework, we implemented the zero range optical limit approximation~(ZROLA) with a nucleon--nucleon~(NN) total cross section that takes the Fermi motion effect into account. As shown in Figs.~\ref{cccs_exp} and \ref{RCS12C12C}, which will be discussed later, the applied Glauber model can reproduce the experimental values of both $\sigma_\mathrm{R}$ and $\sigma_\mathrm{CC}$ simultaneously and consistently.

In this framework, $\sigma_\mathrm{R}$ is expressed using the transmission function $T(b)$ as follows:
\begin{equation}
\sigma_\mathrm{R} = 2\pi \int b \left[ 1 - T(b) \right] db,
\label{rcs}
\end{equation}
\begin{equation}
T(b) \equiv \exp \left[ - \int ds \left(\sum_{i,j} \sigma_{ij}(\bm{b},\bm{s}) \cdot \overline{\rho}_{i}^\mathrm{P}(s) \overline{\rho}_{j}^\mathrm{T}(|\bm{b}-\bm{s}|)\right) \right],
\label{T(b)}
\end{equation}
where $b$ is an impact parameter; $s$ and $|\bm{b}-\bm{s}|$ are the distances from the centers of the projectile and target nuclei, respectively; the indexes $i,j$ denote the isospins of the nucleons in projectile and target nuclei, respectively; $\sigma_{ij}$ is the NN total cross section; $\overline{\rho}^\mathrm{P}(r)$ and $\overline{\rho}^\mathrm{T}(r)$ are density distributions of the projectile and target integrated along beam axis~$z$, i.e., $\overline{\rho}(\bm{b})=\int \rho(\bm{b},z)dz$, respectively.

We applied the effective NN cross section~($\sigma_\mathrm{NN}^\mathrm{eff}$) that includes the Fermi motion effect of the nucleons in the nucleus~\cite{TA09}. Here, $\sigma_\mathrm{NN}^\mathrm{eff}$ was calculated from the bare NN cross section~($\sigma_\mathrm{NN}^\mathrm{bare}$) averaged by the distribution of the relative momentum~($p_\mathrm{rel}$) of the colliding nucleons in the projectile and target nuclei, $D(p_\mathrm{rel})$:
\begin{equation}
 \sigma_{ij}^\mathrm{eff}(\bm{b},\bm{s}) = \int_{-\infty}^{\infty} dp_\mathrm{rel} \sigma_{ij}^\mathrm{bare}(p_\mathrm{rel})D_{ij}(p_\mathrm{rel},\bm{b},\bm{s}),
\label{effnncs}
\end{equation}
\begin{equation}
 \begin{split}
  D_{ij}(p_\mathrm{rel},\bm{b},\bm{s}) = &\cfrac{1}{\sqrt{2\pi\left[ \langle p^2 \rangle^\mathrm{P}_{i}(s) + \langle p^2 \rangle^\mathrm{T}_{j}(|\bm{b}-\bm{s}|)\right]} }\\
  &\times \exp\left[ -\cfrac{\left(p_\mathrm{rel} - p_\mathrm{P}\right)^2}{2\left( \langle p^2 \rangle^\mathrm{P}_{i}(s) + \langle p^2 \rangle^\mathrm{T}_{j}(|\bm{b}-\bm{s}|)\right)} \right],
 \end{split}
 \label{relmomentum}
\end{equation}
where $\langle p^2 \rangle^\mathrm{P}$ and $\langle p^2 \rangle^\mathrm{T}$ are the mean-square momenta of the nucleon in the projectile and target nuclei, respectively, and $p_\mathrm{P}$ is the momentum of the entire projectile nucleus. 
In Ref.~\cite{TA09}, a fixed value of $90$~MeV/$c$ was generally adopted as the value of $\langle p^2 \rangle^{1/2}$ based on the Goldhaber model~\cite{GO74}, while here the $\langle p^2 \rangle^{1/2}$ was calculated from the density-dependent Fermi momentum averaged along $z$:
\begin{equation}
 p_{\mathrm{Fermi},i}(s,z) = \hbar \left[3 \pi^2 \rho_{i}(s,z) \right]^{1/3},
\end{equation}
\begin{equation}
 p_{\mathrm{Fermi},i}^{z}(s) = \cfrac{\int p_{\mathrm{Fermi},i}(s,z)\cdot\rho_{i}(\sqrt{s^2+z^2}) dz}{\int \rho_{i}(\sqrt{s^2+z^2}) dz},
 \label{zaverageFermi}
\end{equation}
\begin{equation}
 \left[\langle p^2 \rangle_{i}(s)\right]^{1/2} = C_\mathrm{FM} \cdot p_{\mathrm{Fermi},i}^{z}(s).
\label{Fermimomentum}
\end{equation}
Here, $C_\mathrm{FM}$ is a constant parameter, which was set to 0.53 to reproduce the energy dependence of experimental $\sigma_\mathrm{R}$ data for $^{12}$C on $^{12}$C~\cite{OZ01,OZ01b,TA09} (Fig.~\ref{RCS12C12C}). The point-proton and point-neutron density distributions $\rho_\mathrm{p}(r)$ and $\rho_\mathrm{n}(r)$, respectively, introduced in Ref.~\cite{TA09}, were used as the density profile of $^{12}$C. Note that $C_\mathrm{FM}=0.53$ is roughly consistent with the Goldhaber model~($C_\mathrm{FM}=0.45$).

In analogy with the Glauber model for $\sigma_\mathrm{R}$, $\sigma_\mathrm{CC}$ is usually formulated by ignoring the contribution of the neutrons in the projectile nucleus. Here, $T(b)$ can be explicitly written according to the projectile composition~\cite{BH04}:
\begin{equation}
 T(b) = T_\mathrm{p}(b)T_\mathrm{n}(b),
\label{TpTn}
\end{equation}
\begin{equation}
T_\mathrm{p}(b) = \exp \left[ - \int ds \overline{\rho}_\mathrm{p}^\mathrm{P} \left\{\sigma_\mathrm{pp}\overline{\rho}_\mathrm{p}^\mathrm{T} + \sigma_\mathrm{pn}\overline{\rho}_\mathrm{n}^\mathrm{T} \right\}\right],
\label{eqTp}
\end{equation}
\begin{equation}
T_\mathrm{n}(b) = \exp \left[ - \int ds \overline{\rho}_\mathrm{n}^\mathrm{P} \left\{\sigma_\mathrm{np}\overline{\rho}_\mathrm{p}^\mathrm{T} + \sigma_\mathrm{nn}\overline{\rho}_\mathrm{n}^\mathrm{T} \right\}\right].
\end{equation}

Then, the charge-changing cross section is obtained as
\begin{equation}
 \tilde{\sigma}_\mathrm{CC} = 2\pi \int b \left[ 1 - T_\mathrm{p}(b) \right] db.
\label{cccs_bare}
\end{equation}
Here, we denote this quantity as $\tilde{\sigma}_\mathrm{CC}$. In this calculation, only $\rho_\mathrm{p}(r)$ in the projectile nucleus is assumed to contribute to the charge-changing cross section. Based on Eq. (\ref{cccs_bare}), in the empirical method~\cite{YA10}, $\sigma_\mathrm{CC}$ is expressed by introducing the energy-dependent scaling factor $\varepsilon(E)$:
\begin{equation}
 \sigma_\mathrm{CC} = \varepsilon(E) \tilde{\sigma}_\mathrm{CC}.
\label{cccs_corr}
\end{equation}
At $E=280$~MeV/nucleon, $\varepsilon=1.123$.

For the Glauber-model calculation, we assumed the two-parameter Fermi-type~(2pF) function for $\rho_\mathrm{p}(r)$ of Ca isotopes:
\begin{equation}
 \rho_\mathrm{p}(r) = \cfrac{\rho_\mathrm{p0}}{1+\exp\left(\cfrac{r-C_\mathrm{p}}{a_\mathrm{p}}\right)},
\end{equation}
where $\rho_{p0}$, $C_\mathrm{p}$, and $a_\mathrm{p}$ are the density constant, half-density radius, and diffuseness, respectively. We assumed that the central density $\rho_\mathrm{p}(0)=\rho_\mathrm{p0}/\left[1+\exp\left(-C_\mathrm{p}/a_\mathrm{p}\right) \right]$ was 0.088~fm$^{-3}$, i.e., half that of the central density of the nucleon density distribution adopted in Ref.~\cite{TA20}. The remaining two parameters were determined to satisfy the known $r_\mathrm{p}$~\cite{GA16} and the volume integral $Z=\int \rho_\mathrm{p}(r)d^3r$. The $r_\mathrm{p}$ was obtained from the RMS charge radius $r_\mathrm{ch}$ as follows:
\begin{equation}
 r^2_\mathrm{p} = r^2_\mathrm{ch} - R_\mathrm{p}^2 - \cfrac{N}{Z}R_\mathrm{n}^2 - \cfrac{3\hbar^2}{4m_\mathrm{p}^2 c^2},
\end{equation}
where $R_\mathrm{p}$ and $R_\mathrm{n}$ are the RMS charge radii of the proton and neutron, respectively, and the last term is the Darwin--Foldy term~\cite{PA16,AT21}.

In Fig.~\ref{cccs_exp}(b), the values calculated using Eqs. (\ref{cccs_bare}) and (\ref{cccs_corr}) are indicated by black solid and green dashed lines, respectively. Although both calculations failed to reproduce the $A$ dependence of experimental data, each worked well in a particular region: $\tilde{\sigma}_\mathrm{CC}$ (Eq. (\ref{cccs_bare})) was closer to the experimental values in the neutron-rich region, but the calculation using the correction factor (Eq. (\ref{cccs_corr})) effectively reproduced the experimental $\sigma_\mathrm{CC}$ of the stable nucleus around $^{42}$Ca. This can be attributed to the determination of $\varepsilon(E)$ from the experimental $\sigma_\mathrm{CC}$ for the stable nucleus, $^{28}$Si~\cite{YA10}. Thus, to explain the overall trend of experimental data, mechanisms other than the conventional Glauber model are required.

\subsection{Introduction of charged-particle evaporation induced by neutron-removal reaction}
Strictly speaking, Eq. (\ref{cccs_bare}) is not based on microscopic theory because it is simply obtained from analogy with the Glauber-model calculation for $\sigma_\mathrm{R}$. Using Eq. (\ref{TpTn}), the relationship between $\tilde{\sigma}_\mathrm{CC}$ and $\sigma_\mathrm{R}$ defined by Eqs. (\ref{cccs_bare}) and (\ref{rcs}) is~\cite{BH04}
\begin{equation}
\begin{split}
 \sigma_\mathrm{R}
\equiv&\; 2\pi \int b \left[ 1 - T_\mathrm{p}(b)T_\mathrm{n}(b) \right] db\\
=&\; 2\pi \int b \left[ 1 - T_\mathrm{p}(b) \right] db \\
&+ 2\pi \int b \left[ T_\mathrm{p}(b) \left\{1 - T_\mathrm{n}(b) \right\}\right] db, \\
\equiv&\; \tilde{\sigma}_\mathrm{CC} + \tilde{\sigma}_{\Sigma-x\mathrm{n}},
\end{split}
\label{defxn}
\end{equation}
where $\tilde{\sigma}_{\Sigma-x\mathrm{n}}$ is the total neutron-removal cross section without proton removal from the projectile nucleus. Bhagwat \emph{et al.} previously noted the neutron-removal contribution to the charge-changing cross section, i.e., $\sigma_\mathrm{CC}\neq\tilde{\sigma}_\mathrm{CC}$~\cite{BH04}. For this reason, Yamaguchi \emph{et al.} introduced a correction factor~\cite{YA10}.

To explicitly incorporate the neutron-removal reaction effect in the $\sigma_\mathrm{CC}$ calculation, the abrasion--ablation model~\cite{GA91}, which consists of two processes to produce the reaction fragments, was adopted. The first stage is abrasion, where a prefragment with excitation energy is produced by abrading nucleons from the projectile nucleus. In the subsequent ablation stage, the prefragment is deexcited to the final fragment through light-particle or gamma-ray emission.

Here, the abrasion--ablation model was introduced similarly to Ref.~\cite{SC04}. Hence, $\sigma_\mathrm{CC}$ was defined as:
\begin{equation}
 \sigma_\mathrm{CC} = \tilde{\sigma}_\mathrm{CC} + \sigma_\mathrm{evap},
\label{cccs_ce}
\end{equation}
where $\sigma_\mathrm{evap}$ is the neutron-removal cross section followed by the charged-particle evaporation and was calculated using the contribution probability of the neutron-removal reaction to $\sigma_\mathrm{CC}$, $P_\mathrm{evap}$. Thus,
\begin{equation}
 \sigma_\mathrm{evap} = P_\mathrm{evap} \tilde{\sigma}_{\Sigma-x\mathrm{n}},
\label{math_pevap0}
\end{equation}
\begin{equation}
\begin{split}
P_\mathrm{evap}
&= \sum_{x=1}^{N_\mathrm{P}} p_{-x\mathrm{n}}\\
&= \sum_{x=1}^{N_\mathrm{P}} r_{-x\mathrm{n}} \int_{0}^{\infty} w_{-x\mathrm{n}}(E_\mathrm{ex}) f(E_\mathrm{ex},A_\mathrm{P}-x,Z_\mathrm{P}) dE_\mathrm{ex}.
\end{split}
\label{math_pevap}
\end{equation}
Here, $E_\mathrm{ex}$ is the excitation energy of the prefragment; $A_\mathrm{P}$, $Z_\mathrm{P}$, and $N_\mathrm{P}$ are the projectile-nucleus mass number, atomic number, and neutron number, respectively; $r_{-x\mathrm{n}}$ is the ratio of the $x$-neutron removal cross section to that of total neutron removal; $w_{-x\mathrm{n}}(E_\mathrm{ex})$ is the excitation-energy distribution of the prefragment in each channel; and $f(E_\mathrm{ex},A_\mathrm{P}-x,Z_\mathrm{P})$ is the probability of charged-particle evaporation for the prefragment with mass and atomic numbers $A_\mathrm{P}-x$ and $Z_\mathrm{P}$, respectively, with $E_\mathrm{ex}$. In addition, $p_{-x\mathrm{n}}$ is the partial value of $P_\mathrm{evap}$ regarding $x$.

\subsubsection{Abrasion stage}
The partial neutron-removal cross section $\tilde{\sigma}_{-x\mathrm{n}}$ was expressed in the following binomial form, similar to the statistical abrasion model~\cite{GA91,BE04}:
\begin{equation}
 \tilde{\sigma}_{-x\mathrm{n}} = 2\pi
\begin{pmatrix}
N_\mathrm{P} \\
x \\
\end{pmatrix}
\int b\mbox{ }T_\mathrm{p}(b) \left[t_\mathrm{n}(b)\right]^{N_\mathrm{P}-x} \left[1-t_\mathrm{n}(b)\right]^{x}db,
\label{eq16}
\end{equation}
\begin{equation}
\begin{split}
 t_\mathrm{n}(b) &= \left[T_\mathrm{n}(b)\right]^{1/N_\mathrm{P}}\\
 &= \exp \left[ - \int ds \left(\cfrac{\overline{\rho}_\mathrm{n}^\mathrm{P}}{N_\mathrm{P}}\right) \left\{\sigma_\mathrm{np}\overline{\rho}_\mathrm{p}^\mathrm{T} + \sigma_\mathrm{nn}\overline{\rho}_\mathrm{n}^\mathrm{T} \right\}\right].
\end{split}
\label{eqsingletrans}
\end{equation}
Here, $t_\mathrm{n}(b)$ represents the probability that a single neutron in the projectile transmits the target density. From Eq.~(\ref{eq16}), $\tilde{\sigma}_{\Sigma-x\mathrm{n}}$ and $r_{-x\mathrm{n}}$ were calculated as
\begin{equation}
 \tilde{\sigma}_{\Sigma-x\mathrm{n}} = \sum_{x=0}^{N_\mathrm{P}}\tilde{\sigma}_{-x\mathrm{n}},
\label{tnrcs}
\end{equation}
\begin{equation}
 r_{-x\mathrm{n}} = \cfrac{ \tilde{\sigma}_{-x\mathrm{n}}}{ \tilde{\sigma}_{\Sigma-x\mathrm{n}}}.
\end{equation}
Note that the sum of $\tilde{\sigma}_\mathrm{CC}$~(Eq.~(\ref{cccs_bare})) and $\tilde{\sigma}_{\Sigma-x\mathrm{n}}$~(Eq.~(\ref{tnrcs})) is mathematically equal to $\sigma_\mathrm{R}$~(Eq.~(\ref{rcs})); i.e., Eq.~(\ref{defxn}) is strictly satisfied.

\subsubsection{Prefragment excitation energy}
The Gaimard--Schmidt method~\cite{GA91} was adopted to determine the prefragment excitation-energy distribution. In this method, the excitation energy distribution of the one-hole state via the single-nucleon removal, $g(\epsilon)$, is defined as
\begin{equation}
 g(\epsilon) = \cfrac{2}{E_\mathrm{max}}\left( 1 - \cfrac{\epsilon}{E_\mathrm{max}} \right).
\label{singlehole}
\end{equation}
This is a linear function satisfying the maximum excitation energy $E_\mathrm{max}$, i.e., $g(E_\mathrm{max})=0$ and $\int_{0}^{E_\mathrm{max}} g(\epsilon)d\epsilon=1$. This functional shape corresponds to an approximation of the single-hole state density in the Woods--Saxon potential~\cite{SC82}. The excitation energy distribution $w_{-x\mathrm{n}}(E_\mathrm{ex})$ via the abrasion of $x$ neutrons is obtained from the convolution of $g(\epsilon)$:
\begin{equation}
\begin{split}
 w_{-x\mathrm{n}}(E_\mathrm{ex}) =& \cfrac{1}{x!}\int_{0}^{\infty} d\epsilon_1  \cdots d\epsilon_{x} \\
&\times \left[g(\epsilon_1)  \cdots g(\epsilon_{x}) \times \delta \left( E_\mathrm{ex} - \sum_{i=1}^{x} \epsilon_{i} \right) \right].
\end{split}
\end{equation}
For the $\sigma_\mathrm{CC}$ calculation within this framework, the only free parameter is $E_\mathrm{max}$.

\begin{figure*}[t]
\resizebox{0.7\textwidth}{!}{\includegraphics{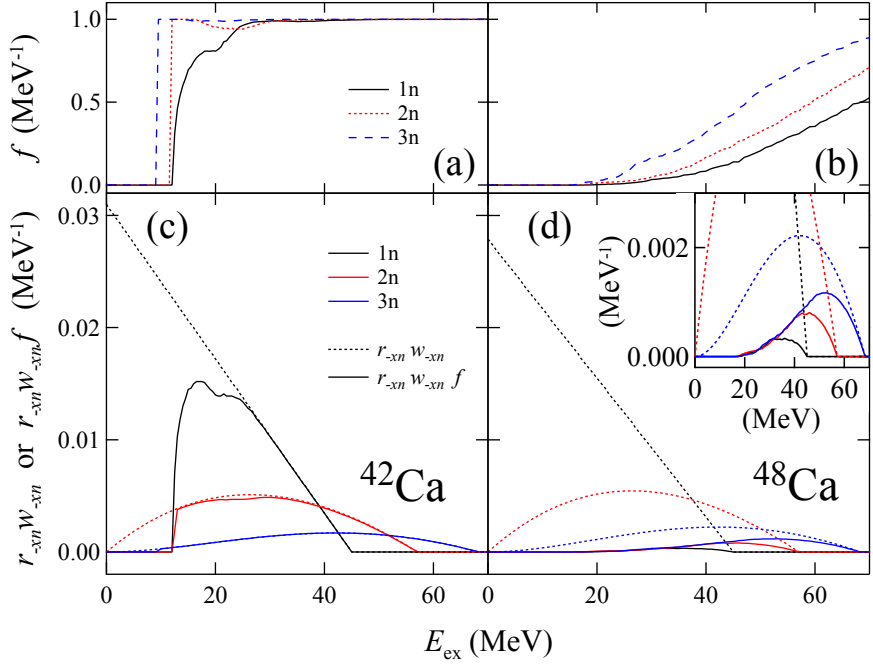}}
 \caption{(a, b) $f(E_\mathrm{ex},A_\mathrm{P}-x,Z_\mathrm{P})$ of 1n~(black), 2n~(red), and 3n~(blue) channels for $^{42}$Ca and $^{48}$Ca. (c, d) $E_\mathrm{ex}$ dependence of $r_{-x\mathrm{n}}w_{-x\mathrm{n}}(E_\mathrm{ex})$~(dotted) and $r_{-x\mathrm{n}}w_{-x\mathrm{n}}(E_\mathrm{ex})f(E_\mathrm{ex},A_\mathrm{P}-x,Z_\mathrm{P})$~(solid) of the respective channels, where $E_\mathrm{max}=45$~MeV. (d, inset) Magnified view. }
\label{Pevap}
\end{figure*}
\begin{figure}[t]
\resizebox{0.5\textwidth}{!}{\includegraphics{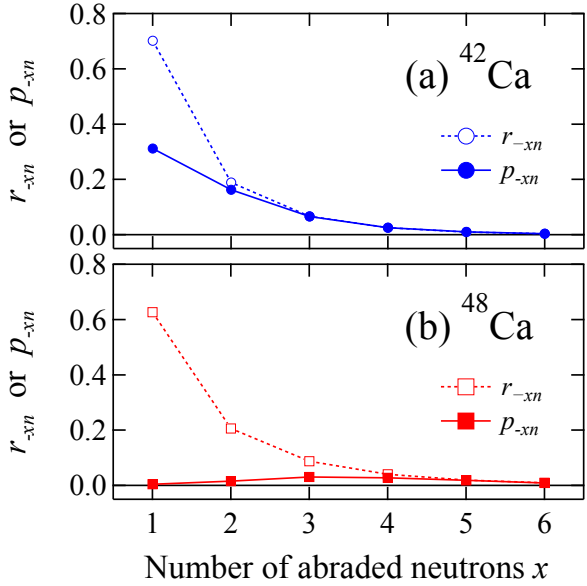}}
\caption{$r_{-x\mathrm{n}}$~(open) and $p_{-x\mathrm{n}}$~(closed) of (a)~$^{42}$Ca~(blue) and (b)~$^{48}$Ca~(red) as functions of the number of removed neutrons~$x$.}
\label{Pevap2}
\end{figure}
\subsubsection{Ablation~(evaporation) stage}
The prefragment deexcitation process was treated using the statistical model, with $f(E_\mathrm{ex},A_\mathrm{P}-x,Z_\mathrm{P})$ calculated using the GEMINI++ code~\cite{CH10,MA10,gemini}. This calculation reflects the light-particle or gamma-ray evaporation from the prefragment nucleus based on the Hauser--Feshbach theory. The yield distribution of the final fragment via the prefragment-nucleus decay was calculated using the Monte Carlo method for all sequential decays until no further decays occurred. Hence, we obtained $f(E_\mathrm{ex},A_\mathrm{P}-x,Z_\mathrm{P})$ from the fraction of all decay channels with at least one charged-particle emission in the sequential decay. Here, the average angular momentum of the prefragment due to the peripheral fragmentation reaction was estimated to be at most $3\hbar$, even for a multi-neutron abraded channel (e.g., four neutrons) of $^{48}$Ca~\cite{JO97}. For such a small angular momentum, $f(E_\mathrm{ex},A_\mathrm{P}-x,Z_\mathrm{P})$ is largely independent on the initial angular momentum; therefore, the initial angular momentum was assumed to be zero here.

\subsubsection{Contribution of respective xn channels to $P_\mathrm{evap}$}
To illustrate the influence of each of the above components on $P_\mathrm{evap}$, calculations using $E_\mathrm{max}=45$~MeV for $^{42,48}$Ca are discussed as examples. In Fig.~\ref{Pevap}~(c, d), results for $r_{-x\mathrm{n}}w_{-x\mathrm{n}}$ (the integral of which over $E_\mathrm{ex}$ and $x$ is 1) are shown for channels up to 3-neutron removal (dotted lines). The $r_{-x\mathrm{n}}w_{-x\mathrm{n}}$ distributions do not differ significantly between $^{42}$Ca and $^{48}$Ca. In contrast, a significant difference is apparent for $f(E_\mathrm{ex},A_\mathrm{P}-x,Z_\mathrm{P})$ between these nuclides (Fig.~\ref{Pevap}(a, b)). For $^{42}$Ca, $f(E_\mathrm{ex},A_\mathrm{P}-x,Z_\mathrm{P})$ of all channels immediately saturates to 1 beyond the threshold energy, which is the sum of the proton separation energy~$S_\mathrm{p}$ and the Coulomb barrier energy~$E_\mathrm{C}$. However, the respective values of $f(E_\mathrm{ex},A_\mathrm{P}-x,Z_\mathrm{P})$ for $^{48}$Ca are far smaller than those for $^{42}$Ca. Qualitatively, this can be interpreted as reflecting the fact that the ratio of the partial widths of the neutron and proton emissions in a single step $\varGamma_\mathrm{n}/\varGamma_\mathrm{p}$, which is dominant in the small $E_\mathrm{ex}$ region, roughly depends on~\cite{BE98}
\begin{equation}
 \cfrac{\varGamma_\mathrm{n}}{\varGamma_\mathrm{p}} \simeq \cfrac{\exp\left[{2\sqrt{a(E_\mathrm{ex}-S_\mathrm{n})}}\right]}{\exp\left[{2\sqrt{a(E_\mathrm{ex}-S_\mathrm{p}-E_\mathrm{C})}}\right]},
\end{equation}
where $a$ denotes the level density parameter. Thus, the competition between proton and neutron evaporation depends on the difference between $S_\mathrm{n}$ and $S_\mathrm{p}$. This tendency generates a small $f(E_\mathrm{ex},A_\mathrm{P}-x,Z_\mathrm{P})$ in the low-excitation-energy region for neutron-rich nuclides. Thus, for $^{48}$Ca, $r_{-x\mathrm{n}}w_{-x\mathrm{n}}f$ (Fig.~\ref{Pevap}(d), solid lines) is distributed only in the high-excitation-energy region.

To clarify the above, $r_{-x\mathrm{n}}$ and $p_{-x\mathrm{n}}$, which is the integral of $r_{-x\mathrm{n}}w_{-x\mathrm{n}}f$ over $E_\mathrm{ex}$, are plotted against the number of abraded neutrons in Fig.~\ref{Pevap2} (dotted and solid lines, respectively). For $^{42}$Ca, 44\% of the 1n channel (which accounts for approximately 70\% of $\tilde{\sigma}_{\Sigma-x\mathrm{n}}$) contributes to $\sigma_\mathrm{CC}$ with charged particle evaporation~($p_{-1n}/r_{-1n}=0.31/0.70=0.44$). Most channels with more abraded neutrons contribute to $\sigma_\mathrm{CC}$, yielding $P_\mathrm{evap}\equiv\sum p_{-x\mathrm{n}} = 0.58$. In contrast, for $^{48}$Ca, $p_{-x\mathrm{n}}$ is almost zero in channels below 3n, yielding $P_\mathrm{evap}=0.11$. Thus, $P_\mathrm{evap}$ strongly depends on the neutron number of the projectile nucleus.

\begin{figure*}[t]
\resizebox{1.0\textwidth}{!}{\includegraphics{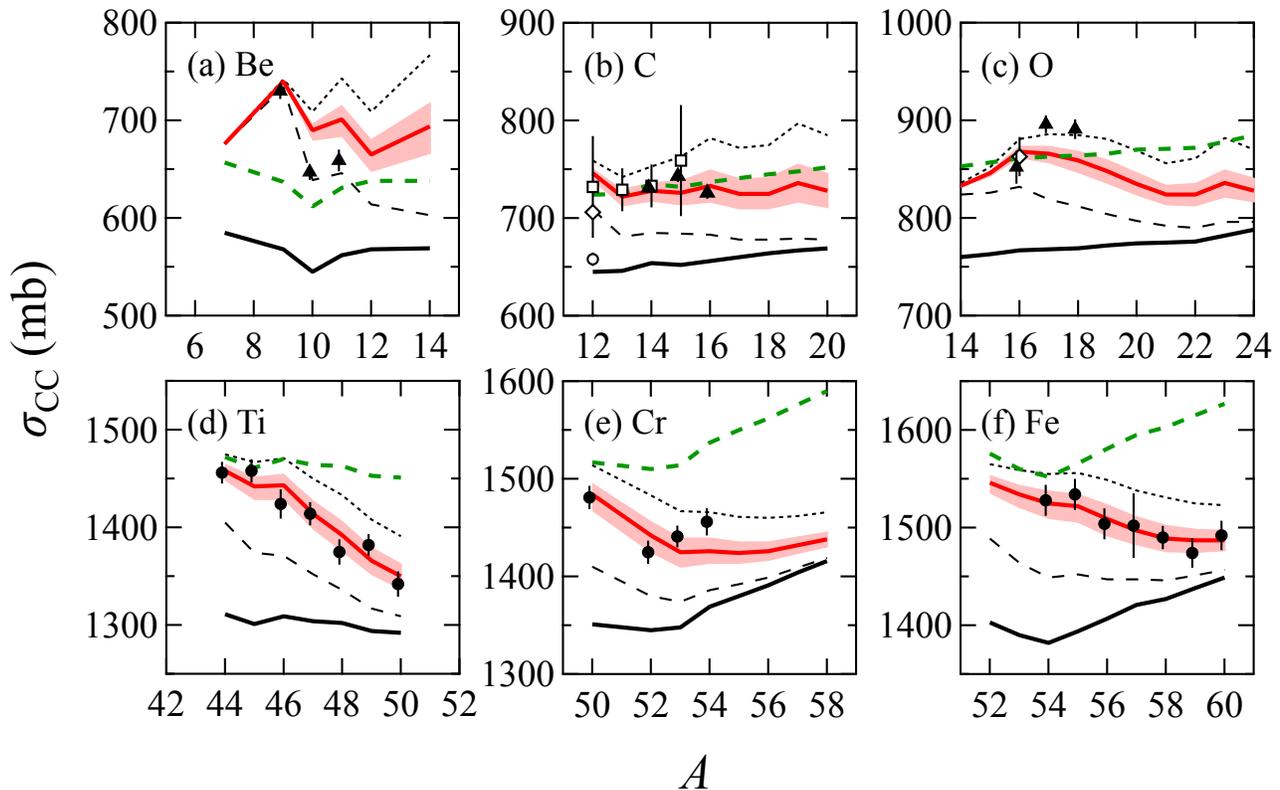}}
\caption{Experimental $\sigma_\mathrm{CC}$ results at $E \simeq 280$~MeV/nucleon and Glauber(CE) $\sigma_\mathrm{CC}$ values for (a)~Be, (b)~C, (c)~O, (d)~Ti, (e)~Cr, and (f)~Fe isotopes as functions of $A$. The experimental results were taken from Refs.~\cite{YA11}~(closed triangles), \cite{YA13}~(closed circles), \cite{ZH20}~(open squares), \cite{ZE07,ZE11}~(open diamonds), and \cite{WE90}~(open circles). Each line is defined as in Fig.~\ref{cccs_exp}.}
\label{cccs_300MeV}
\end{figure*}
\subsection{Comparison of experimental and calculated results}
Hereafter, the calculation introducing the effects described in the previous subsection is denoted ``Glauber(CE).'' To perform this calculation, a 2pF-type function was assumed for the Ca-isotope $\rho_\mathrm{n}(r)$ used in Eq.~(\ref{eq16}). Similar to $\rho_\mathrm{p}(r)$, $\rho_\mathrm{n}(0)=0.088$~fm$^{-3}$ was assumed. Under this constraint, for $^{42\textrm{--}51}$Ca, we employed a 2pF function that reproduced the experimental $\sigma_\mathrm{I}$~\cite{TA20}. For $^{52\textrm{--}54}$Ca, $\sigma_\mathrm{I}$ values extrapolated from $^{48\textrm{--}51}$Ca were used. For $^{40}$Ca, the parameters of the 2pF function were determined to reproduce the experimental $r_\mathrm{m}$~\cite{ZE18}. For $^{36\textrm{--}39}$Ca, the theoretical $r_\mathrm{m}$ values from the Hartree--Fock--Bogoliubov calculation with M3Y-P6a~\cite{NA19} were used to determine the parameters of the 2pF function.

In Fig.~\ref{cccs_exp}(b), the Glauber(CE) results for $\sigma_\mathrm{CC}$ on $^{12}$C for Ca isotopes with $E_\mathrm{max}=20$ and 70~MeV are represented by dashed and dotted thin black lines, respectively. These calculated results depend on the adopted value of $E_\mathrm{max}$. From the chi-square fitting of the Glauber(CE) calculations to the present experimental $\sigma_\mathrm{CC}$ results for $^{42\textrm{--}51}$Ca, $E_\mathrm{max}=45(8)$~MeV was obtained~(Fig.~\ref{cccs_exp}(b), red line). The corresponding $P_\mathrm{evap}$~(Fig.~\ref{cccs_exp}(a)) approaches zero asymptotically in the neutron-rich region, agreeing with the Glauber(ZROLA) calculation for $\sigma_\mathrm{CC}$~(black solid line).

The obtained $E_\mathrm{max}=45(8)$~MeV can be understood by considering a naive Fermi gas model~\cite{GA91}, where the typical Fermi energy is approximately 40 MeV. The $E_\mathrm{max}$ should be obtained when the single hole is located at the potential depth. Under this condition, the single-hole-state energy corresponds to the Fermi energy.

To examine this model for other isotopic chains, we compared the Glauber(CE) results with experimental values of $\sigma_\mathrm{CC}$ on $^{12}$C at around 280~MeV/nucleon for Be~\cite{YA11}, C~\cite{WE90,ZE07,YA11,ZH20}, O~\cite{YA11,ZE11}, Ti~\cite{YA13}, Cr~\cite{YA13}, and Fe~\cite{YA13} isotopes. Here, $\rho_\mathrm{p}(r)$ and $\rho_\mathrm{n}(r)$ were assumed to be the 2pF functions for the Ti, Cr, and Fe isotopes. Harmonic-oscillator-type~(HO) functions~\cite{TA10a} were assumed for the Be, C, and O isotopes:
\begin{equation}
 \rho(r) = \rho(0) \times \left[ 1 + \cfrac{C-2}{3}\left(\cfrac{r}{w}\right)^2 \right]\exp\left[-\left(\cfrac{r}{w}\right)^2\right],
\end{equation}
where $C$ denotes the number of neutrons or protons, $w$ is the radius parameter, and $\rho(0)$ is the normalization factor determined by the volume integral. For nuclides whose experimental $r_\mathrm{p}$~\cite{AN13,MI16} and $\sigma_\mathrm{I}$~\cite{OZ01} results were available, the HO or 2pF function parameters reproducing these results were adopted, as for $^{42\textrm{--}51}$Ca. For $^{14}$Be and the unstable C and O isotopes, no experimental $r_\mathrm{p}$ values have been determined other than from $\sigma_\mathrm{CC}$; thus, the $r_\mathrm{p}$ values were taken from theoretical values via the fermionic molecular dynamics~(FMD)~\cite{TE14}, the coupled-cluster method with NNLO$_\mathrm{sat}$~\cite{TR18}, and relativistic mean field~(RMF) calculations~\cite{TR18}, which reproduce the experimental $r_\mathrm{p}$ values of stable nuclides in their respective isotopic chains~\cite{AN13}. In contrast, $r_\mathrm{p}$ for $^{55\textrm{--}58}$Cr and $^{55,59,60}$Fe were interpolated or extrapolated from experimental $r_\mathrm{p}$ for the corresponding isotopes with $N>28$. This is because $r_\mathrm{p}$ increases linearly in all isotopic chains beyond $N=28$~\cite{MI16}. The parameters for $\rho_\mathrm{n}(r)$ of the Ti, Cr, and Fe isotopes were selected to reproduce the theoretical $r_\mathrm{m}$ obtained via the Hartree--Fock--plus--BCS~(HFBCS) calculations with the SkM* parameters~\cite{INPACS,EB12,EB14}.

The $A$ dependences of $\sigma_\mathrm{CC}$ for respective isotopic chains at around 280~MeV/nucleon are shown in Fig.~\ref{cccs_300MeV}. In the figure, the experimental values decrease with increasing $A$ for the Ca, Ti, Cr, and Fe isotopes, but are rather flat for the Be, C, and O isotopes. Notably, the Glauber(CE) calculations with $E_\mathrm{max}=45$~MeV simultaneously reproduced the experimental results for all above isotopic chains. Existing $\sigma_\mathrm{CC}$ calculation models~(black-solid and green-dashed lines) cannot explain the experimental values for nuclides such as $^{9}$Be, which are always accompanied by charged particle emissions via $^{8}$Be according to the one-neutron removal. However, the Glauber(CE) method successfully reproduced even the experimental value of $^{9}$Be without any special treatment.

\begin{figure}[t]
\resizebox{0.5\textwidth}{!}{\includegraphics{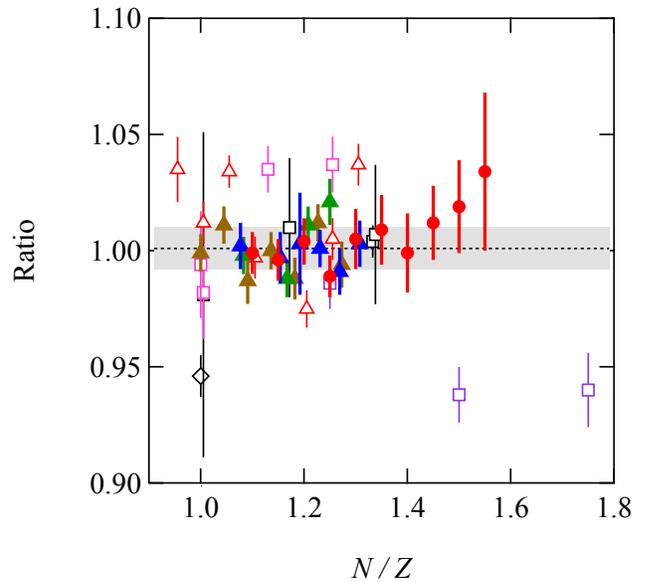}}
\caption{Ratios of experimental $\sigma_\mathrm{CC}$ results on $^{12}$C at around 280~MeV/nucleon to Glauber(CE) calculations for nuclides with previously measured $r_\mathrm{p}$. The open squares indicate the Be~(purple)~\cite{YA11}, C~(black)~\cite{ZH20,YA11} and O~(magenta)~\cite{ZE11,YA11} isotope values. The open diamond indicates the $^{12}$C value from Ref.~\cite{ZE07}. The triangles indicate those of the Ca~(red), Ti~(brown), Cr~(green), and Fe~(blue) isotopes~\cite{YA13}. The present $\sigma_\mathrm{CC}$ results for the Ca isotopes are represented by red closed circles. The shaded band shows the standard deviation~($\pm 0.009$) of the closed symbols around $\mathrm{ratio}=1.001$.}
\label{cccs_residue}
\end{figure}

To quantitatively evaluate the Glauber(CE) method for nuclides with known $r_\mathrm{p}$, the ratios of the experimental $\sigma_\mathrm{CC}$ values to those of the Glauber(CE) calculations with $E_\mathrm{max}=45$~MeV were plotted (Fig.~\ref{cccs_residue}). The Glauber(CE) calculation reproduces the experimental data quite well. The calculations agreed with the experimental values within 1.6\% precision, except for Be isotopes~(purple open squares) and some $^{12}$C data~(black open diamond). In particular, the standard deviation was 0.9\%~(shaded band), and the mean ratio was 1.001 for the Ca, Ti, Cr, and Fe isotopes (Fig.~\ref{cccs_residue}, closed symbols). Thus, the Glauber(CE) calculation improves upon the model ambiguity compared to the previously developed method~\cite{YA11}. This result indicates that $r_\mathrm{p}$ can be determined from $\sigma_\mathrm{CC}$ on $^{12}$C at $E\simeq280$~MeV/nucleon with a systematic uncertainty of approximately 0.5\%, which is comparable to the statistical uncertainty of the typical $\sigma_\mathrm{CC}$ measurement. Because $P_\mathrm{evap}$ is almost zero in the neutron-rich region, the model ambiguity is expected to decrease considerably for neutron-rich isotopes.

\begin{figure*}[t]
\resizebox{1.0\textwidth}{!}{\includegraphics{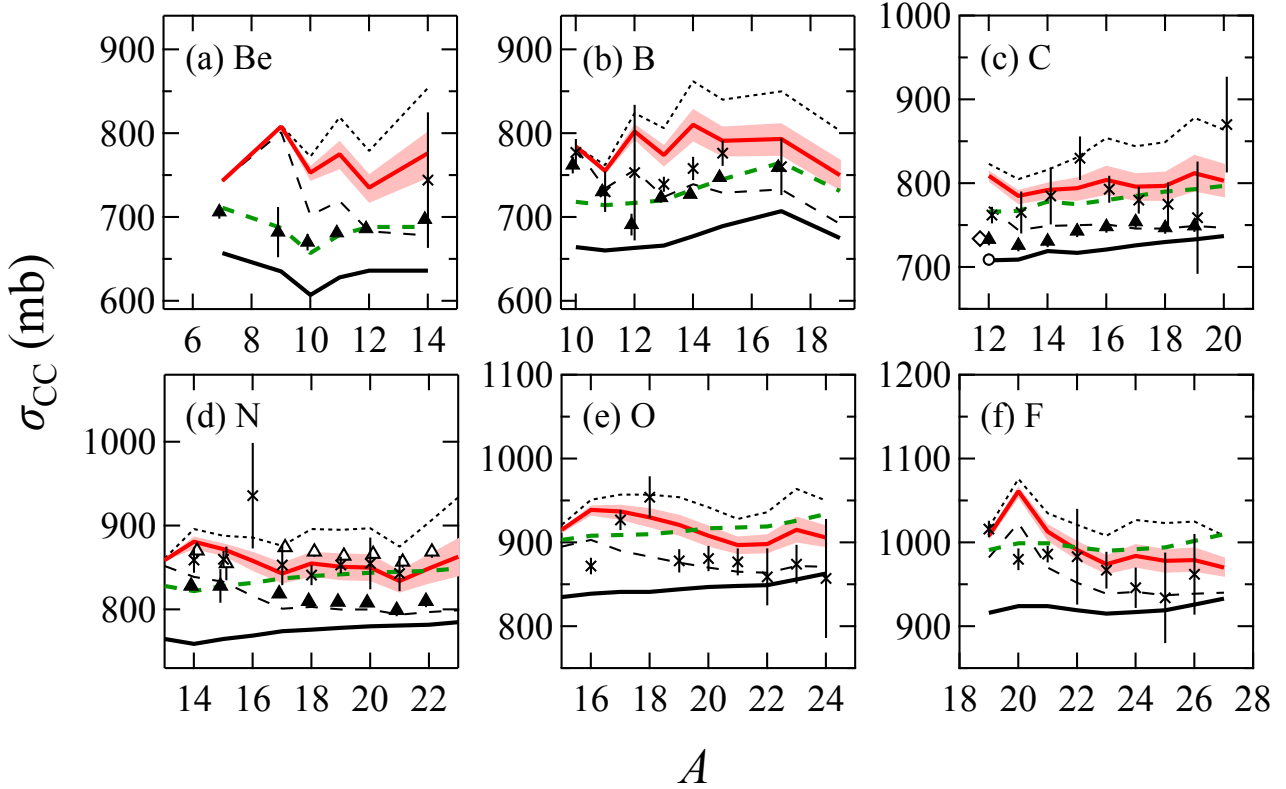}}
\caption{Experimental $\sigma_\mathrm{CC}$ results at $E\simeq900$~MeV/nucleon and Glauber(CE) values of $\sigma_\mathrm{CC}$ for (a)~Be, (b)~B, (c)~C, (d)~N, (e)~O, and (f)~F isotopes as functions of $A$. The systematic experimental results are represented by crosses~\cite{CH00} and closed triangles~\cite{TE14,ES14,KA16,BA19}. The open triangles in (d) indicate $\sigma_\mathrm{CC}^{ex,noveto}$ values from Ref. \cite{BA19}. The open diamond and circle indicate experimental values from Refs.~\cite{TE14} and \cite{WE90}, respectively. Each line is defined as in Fig.~\ref{cccs_exp}.}
\label{cccs_900MeV}
\end{figure*}
\begin{figure}[t]
\resizebox{0.5\textwidth}{!}{\includegraphics{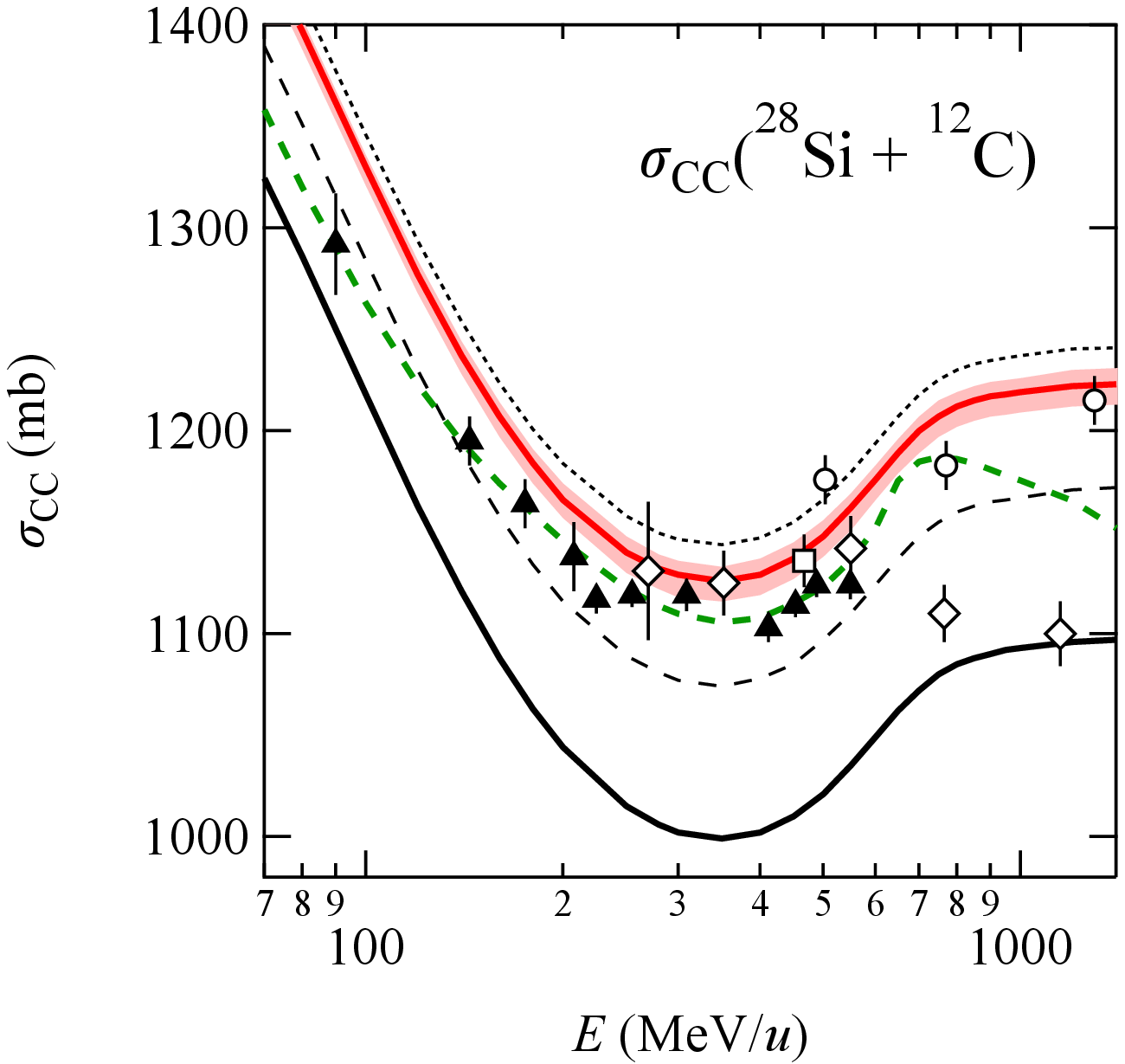}}
\caption{Energy dependence of $\sigma_\mathrm{CC}$ for $^{28}$Si on $^{12}$C. The experimental values are represented by closed triangles~\cite{YA10}, open circles~\cite{WE90}, open diamonds~\cite{ZE07b}, and open squares~\cite{FL01}. Each line is defined as in Fig.~\ref{cccs_exp}.}
\label{cccs_edep}
\end{figure}
The experimental $\sigma_\mathrm{CC}$ values on $^{12}$C at around 900~MeV/nucleon \cite{WE90,CH00,TE14,ES14,KA16,BA19} were also compared. As for the C isotopes above, the $\rho_\mathrm{p}(r)$ and $\rho_\mathrm{n}(r)$ of the B, N, and F isotopes were assumed to be the HO functions that reproduce the theoretical $r_\mathrm{p}$ and experimental $\sigma_\mathrm{I}$~\cite{OZ01,TA17}. As $r_\mathrm{p}$ values to be reproduced, the anti-symmetrized molecular dynamics~(AMD)~\cite{EN15} theoretical values scaled to fit the experimental $r_\mathrm{p}$ of $^{11}$B~\cite{AN13} were used for the B isotopes, the theoretical values from the in-medium similarity renormalization group~(VS-IMSRG)~\cite{BA19}, which reproduce the experimental $r_\mathrm{p}$ of $^{14}$N~\cite{AN13}, were used for the N isotopes, and the theoretical values from the HFBCS with the SkM*~\cite{INPACS,EB12,EB14}, which also reproduce the experimental $r_\mathrm{p}$ of $^{19}$F~\cite{AN13}, were used for the F isotopes.

The experimental and calculated values are presented in Fig.~\ref{cccs_900MeV}. Experimental $\sigma_\mathrm{CC}$ results without the correction of the neutron-removal cross section were taken from Ref.~\cite{ES14,KA16,BA19}, except for those for $^{7,9}$Be~\cite{TE14}. In Ref.~\cite{BA19}, two types of experimental values that depend on the analysis method are reported for N isotopes. Both values, labeled by $\sigma_\mathrm{cc}^{ex,veto}$ and $\sigma_\mathrm{cc}^{ex,noveto}$ in Ref.~\cite{BA19}, are plotted as solid and open triangles in Fig.~\ref{cccs_900MeV}(d), respectively.
The Glauber(CE) calculations with $E_\mathrm{max}=45$~MeV overestimated several values of the Be, B, C, and N isotopes shown by the closed triangles~\cite{TE14,ES14,KA16,BA19}, open circle~\cite{TE14}, and open diamond~\cite{WE90}, and also the values of O isotopes~(crosses). The Glauber(ZROLA) calculations with the correction factor~(green dashed line) and/or the Glauber(CE) calculations with $E_\mathrm{max}=20$~MeV~(thin black dashed line) were rather consistent with these experimental data. In contrast, the experimental values shown by crosses~\cite{CH00} except for O isotopes and open squares~($\sigma_\mathrm{cc}^{ex,noveto}$~\cite{BA19}) agreed relatively well with the Glauber(CE) calculations under the same $E_\mathrm{max}$ as for the data at 280~MeV/nucleon~(red line).

Finally, the energy dependence of the experimental $\sigma_\mathrm{CC}$ on $^{12}$C for $^{28}$Si~\cite{YA10,WE90,ZE07b,FL01} was compared to the Glauber(CE) results (Fig.~\ref{cccs_edep}), for the same $^{28}$Si density profiles as in Ref.~\cite{YA10}. The Glauber(CE) results with $E_\mathrm{max}=45(8)$~MeV overestimated the experimental results in $E<200$~MeV/nucleon. However, at higher energies, the calculation agreed well with the experimental values.
\section{SUMMARY}
In summary, we performed $\sigma_\mathrm{CC}$ measurements for $^{42\textrm{--}51}$Ca on a carbon target at around 280~MeV/nucleon. The obtained $\sigma_\mathrm{CC}$ results decreased significantly with increasing $A$, differing from the trend for light-mass isotopic chains such as C isotopes. The overall experimental trend could not be explained by conventional Glauber-model calculations. To explain the results of Ca isotopes, the charged-particle evaporation effect induced by the neutron-removal reaction was introduced. From the experimental $\sigma_\mathrm{CC}$ data for $^{42\textrm{--}51}$Ca, the parameter of the newly developed Glauber(CE) model, $E_\mathrm{max}$, was determined to be 45(8)~MeV, and experimental $\sigma_\mathrm{CC}$ values at around 280~MeV/nucleon were successfully explained for other isotopic chains from Be to Fe. The deviation between these experimental results and the calculations was 1.6\% for the C to Fe isotopes and, notably, 0.9\% for isotopes beyond Ca. Furthermore, the applied calculation model worked well for several $\sigma_\mathrm{CC}$ data in other energy regions. Thus, the developed method can systematically explain $\sigma_\mathrm{CC}$ over wide mass and energy regions without any phenomenological corrections. The evaporation effect was also found to be negligible for $\sigma_\mathrm{CC}$ for neutron-rich nuclei. Therefore, the experimental $\sigma_\mathrm{CC}$ exactly probes $r_\mathrm{p}$ and the proposed method allows derivation of the  $r_\mathrm{p}$ of very neutron-rich unstable nuclei; these values are difficult to measure using other experimental methods.

\begin{acknowledgments}
We would like to express our gratitude to the accelerator staff at RIKEN Nishina Center for providing the intense $^{238}$U beam. We are grateful to Dr. R.~J.~Charity for providing the latest version of the GEMINI++ code. Discussions of the statistical decay codes with S. Ebata, C. Ishizuka, and K. Sekizawa are also gratefully acknowledged. The present work was supported in part by Grants-in-Aid from JSPS KAKENHI (Grant Nos.~JP24244024 and ~JP16H03905), and a JSPS Research Fellow Grant (No.~JP15J01446).
\end{acknowledgments}

\nocite{*}


\end{document}